\documentclass[aps,twocolumn,showpacs,superscriptaddress,preprintnumbers,
amsmath,amssymb,floatfix,pre]{revtex4-1}  
\usepackage{graphicx}  
\usepackage{dcolumn}   
\usepackage{bm}        
\usepackage{amsmath}   
\usepackage{amsfonts}
\usepackage{amssymb}
\usepackage{natbib}
\usepackage{mathrsfs}  

\usepackage{color}
\definecolor{darkblue}{rgb}{0,0,0.6}
\definecolor{darkred}{rgb}{0.6,0,0}
\definecolor{linkcolor}{rgb}{0,0,0.6} 

\newcommand{\beq}{\begin{equation}}
\newcommand{\eeq}{\end{equation}}
\newcommand{\ba}{\begin{align}}
\newcommand{\ea}{\end{align}}

\begin{document}

\title{Direct calculation of the critical Casimir force in a binary fluid}
\author{Francesco Puosi}
\affiliation{Universit\'e de Lyon, Laboratoire de Physique, \'Ecole normale sup\'erieure de Lyon, CNRS, UMR5672, 46 All\'ee d'Italie, 69364 Lyon, France}
\author{David Lopes Cardozo}
\affiliation{Universit\'e de Lyon, Laboratoire de Physique, \'Ecole normale sup\'erieure de Lyon, CNRS, UMR5672, 46 All\'ee d'Italie, 69364 Lyon, France}
\author{Sergio Ciliberto}
\affiliation{Universit\'e de Lyon, Laboratoire de Physique, \'Ecole normale sup\'erieure de Lyon, CNRS, UMR5672, 46 All\'ee d'Italie, 69364 Lyon, France}
\author{Peter C. W. Holdsworth}
\affiliation{Universit\'e de Lyon, Laboratoire de Physique, \'Ecole normale sup\'erieure de Lyon, CNRS, UMR5672, 46 All\'ee d'Italie, 69364 Lyon, France}
\email{peter.holdsworth@ens-lyon.fr}
\date{\today}

\begin{abstract}
We show that critical Casimir effects can be accessed through direct simulation of a model binary fluid passing through the demixing transition. We work in the semi grand canonical ensemble, in slab geometry, in which the Casimir force appears as the excess of the generalized pressure, $P_{\bot}-n\mu$. The excesses of the  perpendicular pressure, $P_{\bot}$, and of $n\mu$, are individually of much larger amplitude.
A critical pressure anisotropy is observed between forces parallel and perpendicular to the confinement direction, which collapses onto a universal scaling function closely related to that of the critical Casimir force.

\end{abstract}

\maketitle

The critical Casimir \cite{casimir_attraction_1948,m._e._fisher_phenomenes_1978} effect makes a major contribution to nano scale confinement forces \cite{gambassi_casimir_2009}. Its contribution is felt as a fluid is driven through the critical ordering transition of an internal degree of freedom, such as the lambda transition in liquid helium \cite{ganshin_critical_2006}, or the demixing transition of a binary fluid \cite{fukuto_critical_2005,hertlein_direct_2008}. It has been measured in wetting films as they pass close to the critical end point for ordering on the liquid gas interface\cite{ganshin_critical_2006,fukuto_critical_2005}, or through the fluctuation spectra of colloids as their solvent is driven through the demixing transition \cite{hertlein_direct_2008}. Being a critical phenomenon the Casimir effect is characterized by a universal scaling function \cite{krech_casimir_1994} which can be extracted with accuracy from the relevant lattice spin model, relying on the thermodynamic relationship between generalized force and derivative of the free energy \cite{lopes_cardozo_critical_2014,hucht_aspect-ratio_2011,vasilyev_universal_2009,vasilyev_critical_2013,hasenbusch_thermodynamic_2010}. Universality and thermodynamics allow for this dichotomy, but restricts the simulations to average values of the force. Direct simulation of the Casimir force in a fluid \cite{wilding_effect_1998} or model magnet \cite{dantchev_critical_2004}, is a much tougher problem  but accessing it would open the door to instantaneous measurement, microscopic study of the coupling between density correlations and the critical degrees of freedom, as well as to non-equilibrium effects. 

\begin{figure}[th]
\begin{center}
\includegraphics[width=0.62\columnwidth, clip]{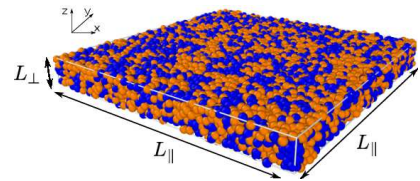}
\end{center}
\caption{Typical configuration for the binary fluid, in slab geometry. Particles of type A (blue), type B (orange). The image is shown for density $n=0.7$ and dimensions $L_{\bot}=5$ and $L_{||}=60$ - see text.}
\label{fig:figure0}
\end{figure}

In this letter we present results showing that universal critical Casimir effects, including a pressure anisotropy and the critical Casimir force, can be accessed through direct simulation of such a binary fluid in an anisotropic cell of volume  $V=L_{\parallel}^2L_{\perp}$, $L_{\bot} \ll L_{\parallel}$ - see Fig.~\ref{fig:figure0}. We study a fluid mixture of species $A$ and $B$ \cite{materniak_phase_2010,das_static_2006,wilding_continuous_2003,wilding_critical_1997} in the semi-Grand Canonical ensemble (SGC) \cite{das_static_2006,materniak_phase_2010}. 
Here one imposes $V$, total number of particles $N=N_A+N_B$, temperature $T$ and chemical potential difference, $\mu_{AB}=(\mu_A-\mu_B)/2$, conjugate to the thermally averaged particle difference, $\Delta N=\langle N_A-N_B\rangle$. The relevant  free energy thus reads $\Omega_{sgc}(T,N,V,\mu_{AB})$. The model system shows a demixing transition characterized by the order parameter $m=\frac{\Delta N}{N}$,  along a line of critical points in the liquid phase,  $(T_C(n),\mu_{AB}=0)$, as the density $n=\frac{N}{V}$ is varied.

The truncation of the diverging correlation length close to the transition introduces $L_{\bot}^{-1}$, measured in microscopic units, as a third variable characterizing the transition, alongside $t=\frac{T-T_C}{T_C}$ and $\tilde{h}=\mu_{AB}/k_BT_C$. The singular dependence on $L_{\bot}$ then leads to the Casimir force \cite{m._e._fisher_phenomenes_1978}. To maintain $T_C(n)$ constant so that $t$ is constant along an isotherme, volume changes must be made at constant density. A similar constraint applies for a magnetic system where volume changes occur in a magnetically homogeneous medium \cite{lopes_cardozo_critical_2014,hucht_aspect-ratio_2011,vasilyev_universal_2009}. Hence, the SGC is thermodynamically equivalent to an Ising model if  volume changes are accompanied by changes in $N$. The fundamental thermodynamic relationship then becomes
\begin{equation}\label{1ere}
d\Omega_{sgc}=-SdT-(P-\mu n)dV-\Delta N d\mu_{AB} \ ,
\end{equation}
with $n$ held constant, where the chemical potential $\mu=(\mu_A+\mu_B)/2$.
From this, one can see that the critical Casimir force should manifest itself in the system size dependence of the generalized pressure, $\tilde{P}=P-\mu n$, conjugate variable to volume changes at fixed $n$, rather than the pressure itself \cite{lopes_cardozo_finite_2015}.

Eqn (\ref{1ere}) must be generalized further to allow for the development of anisotropies when changing  $L_{\perp}$ and $L_{\parallel}$ in slab geometry. This allows independent definitions for the 
pressure measured perpendicular and parallel to the confinement direction \cite{hucht_aspect-ratio_2011,lopes_cardozo_finite_2015}:
\begin{equation}
P_{\perp}=- \left . \frac{1}{L_{\parallel}^2}\frac{\partial \Omega_{sgc}}{\partial L_{\perp}} \right |_{N} , \;
P_{\parallel}=-\left . \frac{1}{2L_{\perp}L_{\parallel}}\frac{\partial \Omega_{sgc}}{\partial L_{\parallel}}\right |_{N},
\end{equation}
and for the generalized pressure $\tilde{P}_{\perp}=P_{\perp}-\mu n$ and $\tilde{P}_{\parallel}=P_{\parallel}-\mu n$.

For a macroscopic sample one finds  $\Omega_{sgc}=k_BT V \omega_{bulk}(T,n,\mu_{AB})$, where $\omega_{bulk}$ is the dimensionless free energy density, so that $P_{\perp}=P_{\parallel}=P=-k_BT \omega_{bulk}$, the bulk pressure. In slab geometry, with $L_{\perp}$ on a meso-scale, corrections to bulk thermodynamics give the announced extra $L_{\perp}$ dependence;  $\Omega_{sgc}=k_BT(\omega_{bulk} + \omega_{ex}(L_\perp))$, where $\omega_{ex}(L_\perp)$ is the excess free energy \cite{goldenfeld_lectures_1992,gambassi_critical_2009,krech_casimir_1994}. Hence one finds an excess values for all thermodynamic quantities including pressure, both perpendicular {\it and} parallel to the confinement direction, internal energy density, $u_{ex}$ and chemical potential, $\mu_{ex}$.

Near the demixing transition there is a critical contribution to the excess free energy, $\omega_{ex}^s$, modifying the singular part of the free energy,
$\omega^s(t,\tilde{h},L_{\perp})=\omega_0^s(t,\tilde{h})+\omega_{ex}^s(t,\tilde{h},L_{\perp})$, where $\omega_0^s$ is the value in the bulk. 

This however, is not the only contribution to the excess. 
Regular surface terms, coming for example, from Van der Waals interactions and leading to a surface free energy \cite{Dzyaloshinskii_general_1961,Cheng_retardation_1988}, also make important contributions that can be taken into account both experimentally \cite{ganshin_critical_2006} and numerically \cite{vasilyev_universal_2009}. In the present work they are minimized as we limit ourselves to periodic boundaries, but even here, the incommensurability of the pair correlation function with the boundaries, leads to non-critical contributions \cite{gonzalez-melchor_stress_2005}. 

The critical excess free energy has universal scaling properties that are at the root of all critical phenomena in a finite system \cite{hasenbusch_specific_2010}. Limiting here to the temperature axis one finds
\begin{equation}
\omega_{ex}^s(t,L_\perp) = L_\perp^{-d} \Theta \left(x_t \right) \; , x_t= t \left( \frac{L_\perp}{\xi_0^+} \right)^{1/\nu} \ .
\label{fully_universal_scaling_omega}
\end{equation}
where $\xi_0^+$ is a non-universal amplitude that depends on $n$, which can be estimated from the small wavelength part of the density structure factor \cite{das_static_2006,das_transport_2003,lopes_cardozo_finite_2015}.

The critical confinement force per unit area, perpendicular (parallel) to the confinement axis is the critical contribution to the excess of $\tilde{P}_{\bot(\parallel)}$
\begin{eqnarray}
f_{\perp}^c &=&( \tilde{P}_{\bot})_{ex}^s/k_BT =  L_\perp^{-d} \theta \left(x_t \right),\nonumber\\
f_{\parallel}^c &=&( \tilde{P}_{\parallel})_{ex}^s/k_BT =-  L_\perp^{-d} \Theta \left(x_t \right),
\end{eqnarray}
with 
\begin{equation}
\theta(x_t)		=	 (d-1) \Theta(x_t) - \left . \frac{x_t}{\nu} \frac{\partial  \Theta}{\partial x_t} \right |_{\tilde{h}} .
\label{theta_Theta}
\end{equation}
The critical Casimir force, in units of $k_BT$ is $f_{\perp}^c$. It indeed depends on the excesses of both the pressure and the chemical potential in this formulation. However, calculation of $\mu_{ex}$ can be bypassed by studying the pressure anisotropy, as the chemical potential is intrinsically isotropic, so that  $\tilde{P}_{\perp}-\tilde{P}_{\parallel}=P_{\bot}-P_{\parallel}$. This is pure excess and the critical contribution to it takes the universal form
\begin{equation}\label{univ-anis}
f^c_{\bot}-f^c_{\|} = L_{\bot}^{-d}\left[ \theta(t L_{\bot}^{1/\nu}) + \Theta(t L_{\bot}^{1/\nu})\right].
\end{equation}
Using Eq.\ref{theta_Theta}, the function $\Theta$ can be calculated from $\theta$ and vice versa   \cite{diehl_large-n_2014}, so that both functions can be extracted with precision from simulations of the Ising model \cite{lopes_cardozo_critical_2014,lopes_cardozo_finite_2015}.

We have tested these ideas using a fluid of A and B particles, interacting via a smoothly truncated potential:
$v(r)=\phi(r)-\phi(r_c)-(r-r_c)\left .\frac{d \phi}{dr}\right |_{r=r_c} \, \text{for } r\leq r_c$
where $\phi(r)$ is the Lennard-Jones (LJ) interaction, $\phi(r)=4\epsilon_{\alpha\beta}[(\sigma_{\alpha\beta}/r)^{12}-(\sigma_{\alpha\beta}/r)^{6}]$ between species $\alpha,\beta$ separated by distance $r$ \cite{allentildesley}. 
We take a symmetric mixture of  equal mass $m$, with $\sigma_{AA}=\sigma_{BB}=\sigma_{AB}=\sigma$, $\epsilon_{AA}=\epsilon_{BB}=2\epsilon_{AB}=\epsilon$, $r_c=2.5\sigma$ and take $\epsilon$, $\sigma$, $m$ and $\tau_0=\sigma\sqrt{(m/\epsilon)}$ as the units of energy, length, mass and time respectively.

We performed hybrid Molecular Dynamics (MD) and Monte Carlo (MC) simulations using a modified LAMMPS code \cite{PlimptonJCP1995}, in slab geometry of width $L_{\|}=60$ and thickness ranging from $L_{\bot}=5$  to $12$ with periodic boundaries (see Fig.~\ref{fig:figure0} for a typical simulation setup). 
The equations of motion were integrated using the velocity Verlet algorithm with a time step $\delta t=10^{-5}$. Constant temperature was achieved using a Nos\'e-Hoover thermostat \cite{NTVnose}. Particle identity switches were made using Metropolis Monte Carlo, with a sweep of attempted changes made every $10^3$ MD steps. At each temperature we started from an equilibrium configuration of $A$ particles and equilibrated for at least $3\times 10^7$ MD steps. 

The pressure $P_k$ in direction $k=x,y$ or $z$ can be accessed by 
the Virial formula \cite{frenkel_understanding_2001},
from which  we define the pressure anisotropy 
\begin{equation}
P_{\bot} - P_{\|} = P_z - \frac{P_x+P_y}{2} \ .
\end{equation}
The interaction part of the chemical potential, over and above that of an ideal gas was calculated using the Widom insertion method \cite{widom_topics_1963,frenkel_understanding_2001}
which was adapted to the SGC with $\mu_{AB}=0$ by inserting either a virtual $A$ or virtual $B$ particle at random. 

The model has been benchmarked in detail for $n=1$ and $\mu_{AB}=0$, showing a second order demixing transition within the dense liquid phase at $T=1.4230\pm 0.0005$ \cite{das_static_2006}.  We anticipate from previous studies of symmetric mixtures, that for this parameter set, 
the phase transition continues along a line of critical points in the $n-T$ plane, intercepting the gas phase at a tri-critical point, \cite{wilding_liquid-vapor_1998} which we avoid by working at sufficiently high densities, $n\ge 0.6$.

\begin{figure}[th]
\begin{center}
\includegraphics[width=0.8\columnwidth, clip]{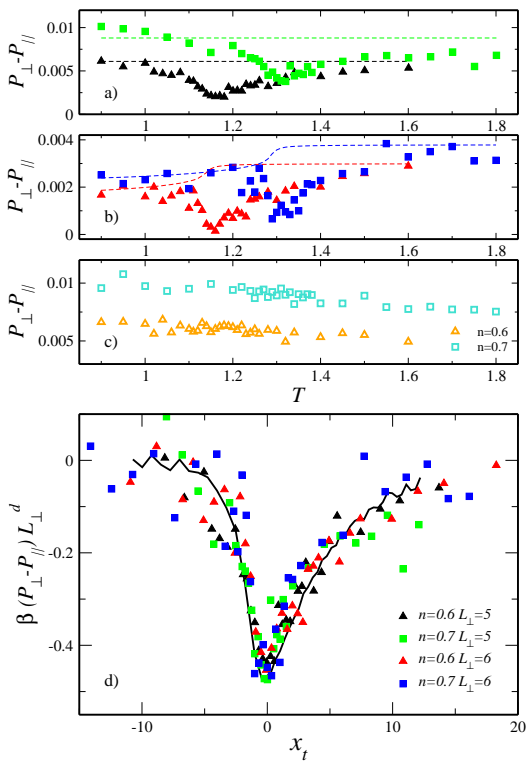}
\end{center}
\caption{{\bf a)} pressure anisotropy $P_\bot - P_{||}$ in slab geometry with $L_\bot=5$ for $n=0.6,0.7$. Dashed lines mark the non-critical pressure anisotropy, estimated from the high and low pressure limits and using eqn. (\ref{shift}) - see text.\\
{\bf b)} pressure anisotropy and estimated non-critical pressure anisotropy for  $L_\bot=6$. \\
{\bf c)} pressure anisotropy for a one component Lennard-Jones fluid for $L_{\bot}=5$ - see text.\\
{\bf d)} finite-size scaling of the critical part of the pressure anisotropy for the systems in panel a) and b). Data are compared to an estimate of the universal scaling function $\theta+\Theta$ computed from a simulation of an Ising model (solid line) with $L_\bot=9$ and $20$ and $L_{||}=60$ \cite{lopes_cardozo_finite_2015}. Color and symbol codes are defined in panels c and d.}
\label{fig:figure2}
\end{figure}

In Fig.~\ref{fig:figure2} we show the pressure anisotropy for $n=0.6$ and $n=0.7$ for $L_\bot=5$ and $L_\bot=6$ in the region of the transition, estimated to be $T_C(n=0.6)=1.18 \pm 0.01$, $T_C(n=0.7)=1.32\pm0.01$. Panels $a$ and $b$ show the raw data as a function of temperature, which illustrate both a critical effect through the transition and a non-critical contribution, resulting in a non-zero value for the anisotropy far from the transition. 

The existence of a non-critical pressure anisotropy for systems confined on this scale \cite{gonzalez-melchor_stress_2005} is illustrated explicitly in panel $c$ where we show data for a single component Lennard-Jones fluid with energy and length scales $\epsilon$ and $\sigma$ for $n=0.6$ in the same temperature range. Away from the transition the anisotropy for the binary and simple fluids are of similar magnitude.
As an ansatz for the non-critical contribution,  to be  subtracted from the total anisotropy we take  
\begin{equation}\label{shift}
P_{nc}(L_{\bot},n,T) = A_{nc}(L_{\bot},n) + B_{nc}(L_{\bot},n) |m|,
\end{equation}
where $m$ is the order parameter for the transition, measured during the simulation and $A_{nc}(L_{\bot},n)$ and $B_{nc}(L_{\bot},n)$ are constants estimated from the high and low temperature values for the data shown in panels $a$ and $b$. This represents the pressure anisotropy of an effective single component fluid whose characteristics evolve as the composition of the binary fluid evolves through the transition. The amplitude of this evolution is fitted, but the rate, as a function of temperature is fixed by the evolution of $m$, which is determined numerically. For $n=0.6$ it was sufficient to set $B_{nc}=0$, giving a constant shift shown by the dashed line in panel $a$. For $n=0.7$ both constants were taken to be non-zero with the resulting shift functions shown in panel $b$. 

Removing the non-critical contribution, we arrive at one of the main results of the letter in panel $d$, which shows the reduced pressure anisotropy plotted against $x_t$. The fluid data collapses, within statistical error onto a single master curve in excellent agreement with that for the Ising model in which the universal function has been constructed from the Casimir force and excess free energy, calculated independently \cite{lopes_cardozo_finite_2015}. This function shows a minimum value just below, but close to the transition, with an overall form  similar to $\theta(x_t)$ for the Casimir force.  
The extra term from the excess free energy generates a broader function, particularly above the transition and  gives a depth at $T=T_C$ which is $3/2$ that of $\theta(x_t)$ \cite{lopes_cardozo_finite_2015}. Given the small values for $L_{\bot}$, one can expect the results to be subject to corrections to scaling \cite{vasilyev_universal_2009,hasenbusch_thermodynamic_2012-1,pelissetto_critical_2002,hasenbusch_finite_2010-1} but these are not resolved for the present data sets.

\begin{figure}[th]
\begin{center}
\includegraphics[width=\columnwidth, clip]{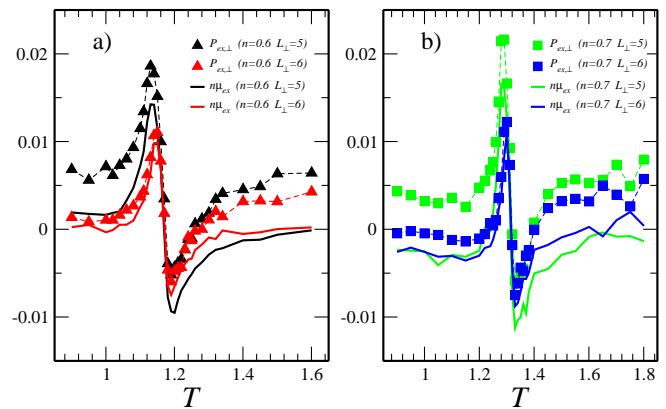}
\end{center}
\caption{Excess part of the confinement pressure $P_{ex,\bot}$ (symbols), and excess chemical potential multiplied by the density, $n\mu_{ex}$ (full lines), for $L_\bot=5,6$ and density $n=0.6$ {\bf (a)} and $n=0.7$ {\bf (b)}.  Colors and symbols defined in the figure. }
\label{fig:figure3}
\end{figure}

In Fig.~\ref{fig:figure3} we show the excess values for the pressure measured perpendicular to the confinement direction, $ P_{ex,\bot} = P_{\bot}  - P^{cubic} $ and for the product of density and chemical potential, $n \mu_{ex} =n \mu - n\mu^{cubic}$. The quantities $P^{cubic}$ and $\mu^{cubic}$ were estimated from a sample with $L_{\bot}=L_{\parallel}=32$ and within the precision of our simulations can be considered as the bulk values \cite{hucht_aspect-ratio_2011,lopes_cardozo_finite_2015}.
The first thing to notice is that these two quantities exhibit a sharply varying alternating function through the transition, similar in form to the excess internal energy \cite{lopes_cardozo_finite_2015,hasenbusch_specific_2010} but radically different from the Casimir function exposed by the pressure anisotropy. Secondly, comparing with Fig.~\ref{fig:figure2} panels $a$ and $b$, one can see that the amplitude of these excess quantities is an order of magnitude bigger than that expected for the Casimir force for $L_{\bot}=5$ and $6$.  These measurements then confirm the thermodynamics presented above: in the SGC ensemble the Casimir force is related to the excess of $\tilde{P}_{\perp}$ rather than the pressure itself. Its direct calculation poses the numerical challenge of resolving this small signal from the difference between these two quantities. 
\begin{figure}[th]
\begin{center}
\includegraphics[width=\columnwidth, clip]{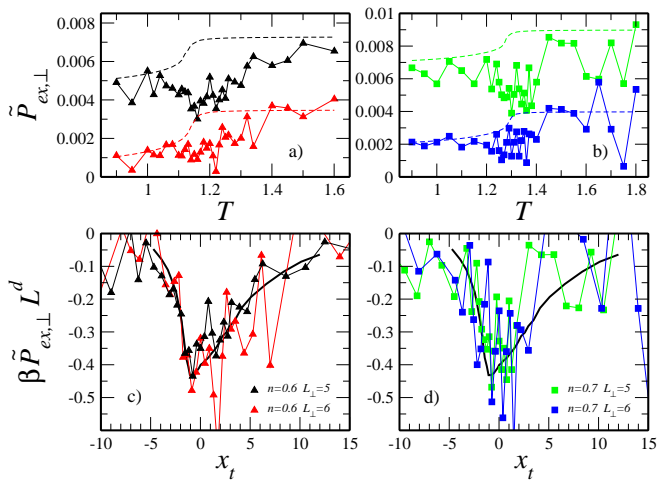}
\end{center}
\caption{{\bf Panel a) and b)}: symbols correspond to the excess part of the generalized pressure $\tilde{P}_{ex,\bot}$ with $L_\bot=5,6$ and for $n=0.6$ (a) and $n=0.7$ (b). Dashed lines are estimates of the non critical part of $\tilde{P}_{ex,\bot}$. {\bf Panel  c) and d)}: scaling of  $\tilde{P}_{ex,\bot}$ the scaling function $\theta(x_t)$ obtained from Ising model simulation in ref. \onlinecite{lopes_cardozo_critical_2014} (full line). Colors and symbols defined in panels c) and d). }
\label{fig:figure4}
\end{figure}

In Fig. \ref{fig:figure4}, panels $a$ and $b$, we show $\tilde{P}_{ex,\bot}$ {\it vs.} $T$ for $L_\bot=5,6$ and $n=0.6$ and $n=0.7$ respectively, constructed from the data shown in Fig.~\ref{fig:figure3}. The raw data shown in panels $a$ and $c$ fluctuates around values which are of the correct order of magnitude for the Casimir force but its observation is obscured by what again we interpret as  a non-critical contribution, $\tilde{P}_{nc}(L_{\bot},n,T)$. We model $\tilde{P}_{nc}$ using the functional form of eqn. (\ref{shift}). The results are superimposed on panels $a$ and $b$.  In each case the constants $A$ and $B$ were fixed by estimating the asymptotes at high and at low temperature. The subsequent data collapse, once $\tilde{P}_{nc}$ has been subtracted, is shown in panels $c$ and $d$, and compared with the universal function, as estimated from simulations of the Ising model \cite{lopes_cardozo_critical_2014}. The data is extremely noisy and the process is clearly somewhat exploratory, but despite this it seems clear that the universal function is emerging from the simulation data for the binary mixture.

In conclusion, we have shown that one can  successfully access Casimir effects directly from numerical simulations of  a binary fluid undergoing a critical demixing transition. We worked here in the SCG ensemble,  both because it is thermodynamically equivalent to the Ising model, and because it offers one of the most accessible routes to direct simulation in a fluid system. We were able to show that the confinement induces a pressure anisotropy in the fluid, with  the pressure components perpendicular and parallel to the confinement direction showing distinct excess contributions coming from the truncated critical fluctuations. Measurement of $P_{\bot}-P_{\parallel}$ yields a universal scaling function intimately related to the critical Casimir force. Direct calculation of the Casimir force requires the calculation of the excess of the generalized pressure  $\tilde{P}_{\bot}=P_{\bot}-n\mu$, and resolving a signal from the difference between these already highly fluctuating quantities proved to be at the limit of our computing resources. In particular, obtaining data for the chemical potential of high enough quality was extremely time consuming and the key to future improvements is the development of more efficient algorithms for estimating $\mu$. 

An alternative to our approach would be to work in the grand canonical ensemble, fixing $\mu$ and allowing fluctuations both of $N$  and  the number of each species, $N_A$ and $N_B$ \cite{wilding_continuous_2003,wilding_liquid-vapor_1998,wilding_effect_1998}.  
In this case, as the chemical potential fixes the average density during a volume change, the Casimir force should come directly from the excess pressure exerted on the confining walls. However, simulating 
density fluctuations is time consuming, making high quality data collection difficult. As a consequence, calculation of the pressure anisotropy and other excess quantities such as the internal energy are more straightforward in the SGC. For the Casimir force, things are less clear and 
in the light of the results presented here it would certainly be interesting to explore grand canonical simulations further.

Making direct simulations of a binary fluid has highlighted the existence of the pressure anisotropy associated with confinement; a rather unusual situation for an otherwise isotropic fluid. This anisotropy could in principle be measured experimentally and we hope that our work will stimulate the development of protocols for this. One possible avenue could be through the evolution of the forces between arrays of colloidal particles pinned in  anisotropic geometries, immersed  in a binary fluid as it passes through the demixing transition.

Finally, the most compelling motivation for making direct simulations in a fluid system, is that it allows the study of both universal properties and microscopic processes. In addition, the use of molecular dynamics gives access to real dynamical processes and will open the door to studies of non-equilibrium processes. 
In this paper, we have been limited to periodic boundaries, but while more realistic boundaries will require increased computing power and efficiency, there is no theoretical barrier to the study of fixed, or open boundaries, corresponding to experimental set ups. Future experiments will certainly move from detailed studies of Casimir forces in equilibrium situations to the study of non-equilibrium phenomena. Numerical simulations of model fluid systems such as those presented here will play an important role in understanding these exciting developments

We thank J.-L. Barrat and A. C. Maggs for useful discussions. This work was financed by the ERC grant OUTEFLUCOP and used the numerical resources of the PSMN at the ENS Lyon. PCWH acknowledges financial support from the Institut Universitaire de France.

\bibliographystyle{unsrt}
\bibliography{./biblio}

\end{document}